\documentclass[3p, preprint, times, lefttitle, twocolumn]{elsarticle}

\usepackage[english]{babel}
\usepackage[utf8]{inputenc}
\usepackage{wrapfig}
\usepackage{graphicx}
\usepackage{amsmath}
\usepackage{multirow}
\usepackage{hyperref}
\hypersetup{
    colorlinks=true,
    linkcolor=black,
    filecolor=magenta,      
    urlcolor=cyan,
    }

\title{Novel Data Analysis Tool for the Evaluation of Coincidence Doppler Broadening Spectra of the Positron-Electron Annihilation Line}

\author[1]{Leon Chryssos}
\ead{leon.chryssos@frm2.tum.de}
\author[1]{Christoph Hugenschmidt\corref{cor1}}
\ead{christoph.hugenschmidt@frm2.tum.de}
\affiliation[1]{organization={Heinz Maier Leibnitz Zentrum (MLZ), Technical University of Munich},
addressline={Lichtenbergstr. 1},
postcode={85748},
city={Garching},
country={Germany}}
\cortext[cor1]{Corresponding author}

\date{November 2022}

\journal{Nuclear Instruments and Methods in Physics Research Section A}

\begin{document}

\begin{abstract}
Coincidence Doppler Broadening Spectroscopy (CDBS) of the 511\,keV  annihilation line reveals the elemental signature at the annihilation site in matter.
For this reason, CDBS enables the analysis of foreign atoms in the host matrix, vacancy-solute complexes and precipitates in solids. 
Due to the lack of a comprehensive program for the analysis of CDBS data we developed a modular python-based open source software solution for the reliable extraction of the annihilation line to provide so-called ratio curves highlighting the elemental signature in the spectra.
In particular, the new program allows the readout and spectra averaging of multiple detector pairs. 
Moreover, we implemented an improved algorithm to further reduce the background by about one order of magnitude at high Doppler shifts.
\end{abstract}

\bibliographystyle{elsarticle-num}

\maketitle

\section{Introduction}
\label{sec:intro}
Positron Annihilation Spectroscopy (PAS) is a well-established technique for the characterization of condensed matter properties such as the concentration and distribution of open volume defects. In matter positrons typically thermalize within a few ps \cite{Krause99} -- i.e. within a time range two orders of magnitude smaller than their typical lifetime of 100\,ps \cite{Sch88}.
Lattice defects like vacancies create attractive potential wells that efficiently trap thermalized positrons. 
Hence, together with an electron the positron annihilates predominantly into two $511 \ \mathrm{keV}$ $\gamma$ quanta either from a delocalized state in the unperturbed lattice or from a trapped state in an open-volume defect. 
The positron momentum is usually negligible since it annihilates in thermal equilibrium; the on average much higher momentum distribution of the annihilating electrons, however, is encoded in the emitted annihilation radiation. Due to momentum conservation the transferred momentum results in an angle deviation from 180$^\mathrm{o}$ and a Doppler shift of the emitted $\gamma$ quanta.  

For this reason, Doppler Broadening Spectroscopy (DBS) of the positron-electron annihilation line provides valuable information of the annihilation site in the sample. In general, preferred annihilation of core electrons with larger momenta than valence electrons will result in a broadening of the 511 keV annihilation photo peak. In vacancies the annihilation of positrons with core electrons is significantly decreased compared to the defect-free lattice. Consequently, vacancy-like lattice defects lead to a narrowing of the annihilation line. 

The 511\,keV annihilation line intrinsically contains the projection of the electron momentum distribution of the investigated sample, weighted by the annihilation probability of the positron with each electron orbital. However, due to the unavoidable Compton and pile-up background most of this information is obscured in DBS spectra. In Coincidence Doppler Broadening Spectroscopy (CDBS) both $\gamma$ quanta are detected simultaneously by using two opposing high-purity Ge detectors. The resulting measurement of the total annihilation energy of $1022 \ \mathrm{keV}$ leads to a significant increase of the signal-to-noise ratio allowing the separation of true annihilation events from the background. 

The low background of CDBS was shown to enable the analysis of open volume defects with superior accuracy compared to conventional DBS \cite{Sei16}. But more importantly, CDBS reveals element information at the annihilation site \cite{Aso96}. For example, it was demonstrated, that the shape of CDB spectra directly correlates with the number of d-electrons of pure elements such as Ti, V, Cr, Fe, Ni and Cu \cite{Bar04}.
Hence, CDBS enables the determination of the chemical surroundings of lattice defects. 
Examples are the investigation of irradiation induced defects in Mg \cite{Sta07a} or FeCr alloys \cite{Par15}, the identification of the species of point defects in MnSi crystals \cite{Rei16a}, or the position resolved study of oxygen vacancies in high-Tc superconductors \cite{Rei18}. 
By calculating CDB spectra in theoretical studies subtle effects have been taken into account such as the vacancy size of vacancy-solute complexes \cite{Kur06}, isotropic inward relaxation of the nearest-neighbours of vacancies \cite{Cal05} or the influence of the number of atoms of different elements decorating a vacancy \cite{Fol07a}.
In this context, it has to be emphasized that clusters, e.g., precipitates, can be identified provided they exhibit a higher positron affinity than the host lattice  \cite{Pus89}. Such studies comprise the analysis of irradiation-induced Cu aggregations in Fe \cite{nag01}, metal layers embedded in Al \cite{Hug08a, Pik11} or position resolved detection of Cu-rich precipitates in welded Al alloys \cite{Gig17, Bac19}. 

In CDBS studies, it has become common to show so-called ratio curves in order to highlight the elemental fingerprint of the annihilation site in a sample. For the generation of ratio curves, the measured 2D-CDB spectra are projected onto one energy axis, normalized and divided by a reference spectrum of interest. The CDB spectrometer of our research group \cite{Gig17} uses the remoderated positron beam \cite{Pio08} available at the NEutron induced POsitron source MUniCh (NEPOMUC) \cite{Hug02a} at the research neutron source Heinz Maier-Leibnitz (FRM II) of the Technical University of Munich (for details of the beam characteristics see \cite{Hug13b, Hug14b}). After a recent upgrade with five detector pairs, the need of a new program for data treatment allowing fast and accurate averaging of CDB spectra became apparent.

The proper data treatment for obtaining reliable and statistically most meaningful 1D energy spectra from CDBS measurements is demanding. 
In order to solve all the issues that come with the data processing of CDB spectra we decided to develop a comprehensive and modular python-based program. 
This program comprises the extraction of the annihilation line by accurately taking into account the background, handling data and averaging spectra from multiple detector pairs as well as the generation and plotting of ratio curves.
Finally, the software package is open-source and hence available to all users in the scientific community.

\section{Coincidence Doppler Broadening Spectroscopy}
\subsection{CDBS versus DBS}
In conventional DBS the energy of just one of the two annihilation quanta is measured using a single Ge detector.
The high Doppler shift regions at the tails of the $511 \ \mathrm{keV}$ annihilation line are therefore obscured by background events.
In order to significantly improve the signal-to-noise ratio at the annihilation line, both $\gamma$-quanta are detected simultaneously in CDBS. 
Since the angle deviation from 180° of the annihilation $\gamma$-quanta is very small, two Ge detectors facing each other with the annihilation site in their center can be used.
A sufficiently small coincidence time window is chosen typically in the range of hundreds of $\mathrm{ns}$ to avoid random coincidences. 

\begin{figure}[h]
    \centering
    \includegraphics[width=\linewidth]{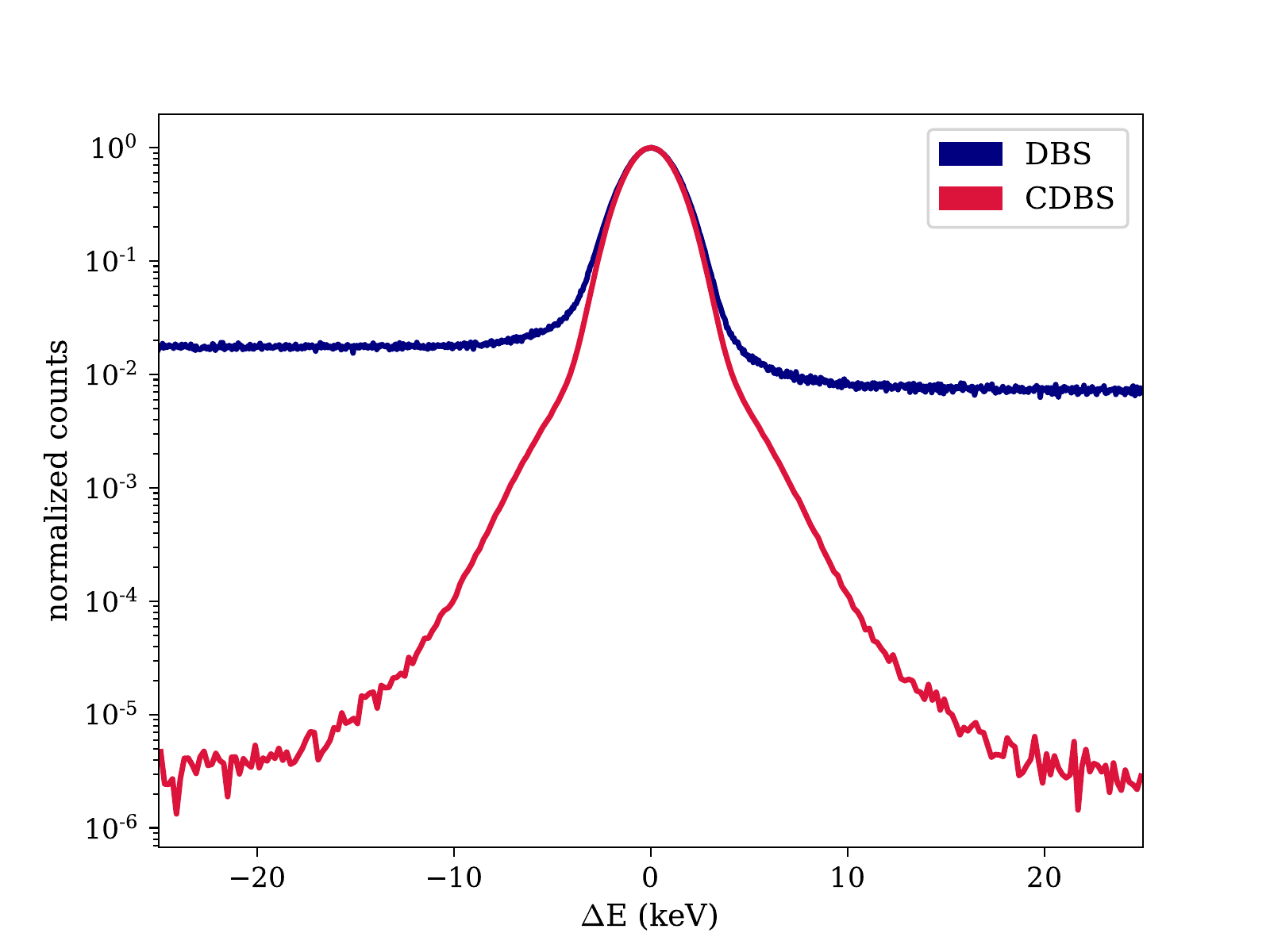}
    \caption{Comparison of DBS and CDBS of the Doppler-broadened annihilation photo peak recorded for Cu. A single detector is used in conventional DBS (blue). The projection of the CDBS measurement (red) is performed by the algorithm from the new software package.}
    \label{fig:BDS_CDBS}
\end{figure}

As demonstrated in Figure\,\ref{fig:BDS_CDBS}, the pileup of (Compton) events around 511\,keV lead to a large background in conventional DBS.
With CDBS the improved signal-to-noise ratio leads to an efficient background reduction of almost four orders of magnitude compared to DBS.
Hence, annihilation events with core electrons are revealed enabling the analysis of element-specific signatures in high Doppler-shift regions of the annihilation line. 

From a technical point of view, it is not trivial to extract the annihilation line from data generated by CDBS while accurately taking into account the background. CDBS generates data in the form of a two-dimensional (2D) histogram wherein the area including the annihilation line does not correspond to the bins generated by the digitization of the detector signals. This means that simply re-binning the 2D histogram will introduce artefacts and significant errors which become especially noticeable at high energy resolutions. Additionally, the data is convolved with a two-dimensional energy resolution function that is specific to the experimental CDBS setup. Due to the noisiness of the data a de-convolution is very difficult if at all possible to perform \cite{ho03}. 

Another approach is to extract the projected annihilation line by successive Gauss fitting \cite{Pikart2014}. By design this method intrinsically suppresses the background of the measurement data. However, the use of Gauss fits to project the annihilation line is prone to convergence errors and hence introduces faulty data especially for measurements with lower statistics. Finally, the data can be extracted by simply defining a rectangular region of interest (ROI) and then re-binning along the annihilation line to project the data. This approach is very robust, however, care has to be taken that when cutting the data an interpolation step has to be performed, otherwise artifacts will be introduced. This can be combined with either a constant background subtraction method or a specially modeled background, fitted to the 2D histogram \cite{step17}.

\begin{figure}[h]
    \centering
    \includegraphics[width=1.1\linewidth]{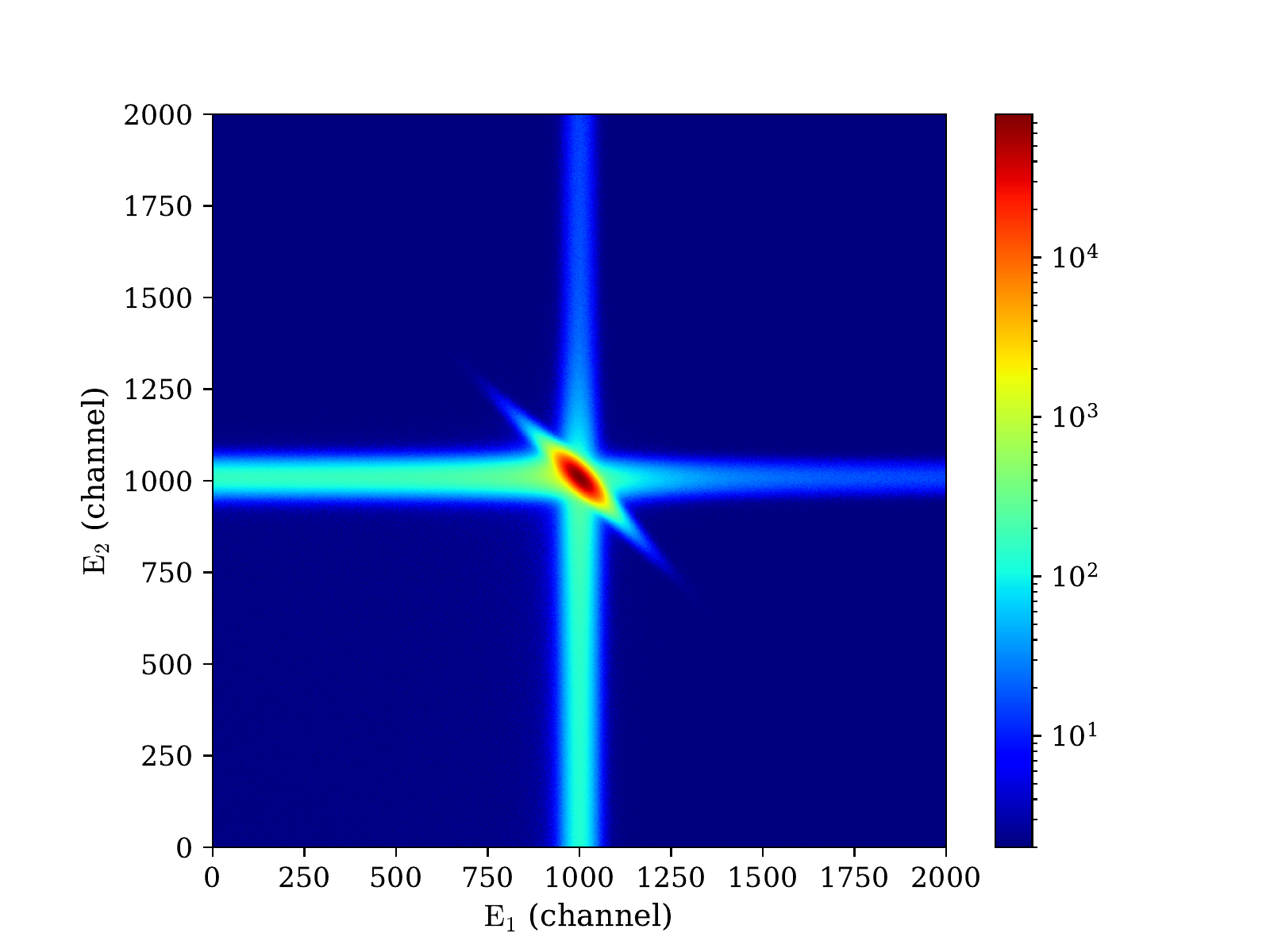}
    \caption{2D histogram of a CDBS measurement on Cu. 
    The energies $E_1$ and $E_2$ refer to the respective detector 1 and 2. 
    The diagonal from top left to bottom right represents the constant total energy of $1022 \ \mathrm{keV}$.
    The cross pattern around the $511 \ \mathrm{keV}$ photo peak in the center is caused by Compton background (on the lower left) and by pileup events (on the top right).}
    \label{fig:hist}
\end{figure}

In CDBS, every annihilation event has two associated energies and is typically stored in a 2D array. 
The 2D matrix containing the coincidence data can be plotted in a 2D histogram as shown in Figure\,\ref{fig:hist}. 
The axes correspond to the $\gamma$ energies $E_1$ and $E_2$ measured by detector one and detector two respectively.
Note that the raw data is stored in channels which are proportional to the detected energy; the correct conversion from \textit{channel} to \textit{keV} is done by using the energy calibration of the detectors as explained in Section\,\ref{sec:roi}.
Using this visualization one can easily identify the advantage of coincidence measurements since the annihilation photo peak is easily distinguishable from background events. In the 2D histogram the Doppler-broadened photo peak appears on a constant energy line (CEL) at $1022 \ \mathrm{keV}$ almost background-free. Beside the Doppler peak a cross shaped pattern appears as a result of coincidences of Compton scattered or pileup events and events in the $511 \ \mathrm{keV}$ photo peak. 

For the accurate quantitative analysis though, separating the photo peak from the background is less straight forward. 
The $1022 \ \mathrm{keV}$ CEL is not only broadened along itself due to Doppler shifts. It is also broadened, both perpendicular and parallel to itself, due to the experimental energy resolution of the detectors.

\subsection{1D- Projection from the 2D-Histogram}
\label{sec:proj}
As mentioned above there are several ways to extract the CEL from the rest of the spectrum, for example, by performing perpendicular Gauss fits at predefined energies, as it was realized by P.\,Pikart  \cite{Pikart2014}. Due to the nature of the Gaussian fits this approach offers excellent background suppression.

The other way to extract the photo peak is defining a ROI with a fixed shape. On its own this method seems to be inferior to the first one due to worse background suppression. This, however, is not necessarily the case since the susceptibility of the ROI method, with regard to background events, can be significantly improved with a simple background subtraction step. In addition, there are no convergence issues with fits providing a more robust data extraction pipeline and lower execution times.

\subsection{Ratio Curves}
Similarly to DBS, the most commonly used analysis method for CDBS are comparisons to reference samples. However, CDBS can not only detect minute variations of defect concentrations, but also differences in the chemical environment of the annihilation site. In order to visualize these differences so-called ratio curves are commonly utilized. A reference sample is selected and subsequently the projections of all measurements of interest are divided by the projection of the reference spectrum.
An example of such ratio curves with annealed Al as a reference is given later in figure\,\ref{fig:ratio}. 

\section{New Analysis Tool for (C)DBS Spectra}
We developed a Software To Analyze CDB Spectra (STACS) due to the lack of a complete and versatile software package for the accurate generation of CDB spectra, i.e., 1D-projections of a well-defined ROI of the recorded 2D-energy spectra.
The availability of such a program is of utmost importance to reliably extract the elemental information gained by CDBS.

The following conditions and features were identified as aims to be fulfilled by STACS:
\begin{itemize}
    \item Use of a modular python-based program structure to run the software stand alone using the users own script and making it easy to adjust and expand. This also enables the integration of  individual objects or methods into other existing codes or user generated scripts.
    \item Implementation of a robust and simple method to extract the annihilation line while negating artefacts that often arise when handling 2D binned data. This allows the reliable generation of the 1D projection without convergence problems or artifacts, independent of statistics.
    \item Application of an accurate background subtraction step --in addition to the intrinsic background suppression of CDBS-- to further increase the signal-to-background ratio.
    \item Calculation and visualisation of ratio curves with great flexibility. The data can also just be exported in order to be used with any other visualization program.
    \item Handling data and averaging spectra from multiple detector pairs in order to improve the statistics. 
    \item  Offering the software to the scientific community as an open source package for free download.
\end{itemize}

STACS consists of two main clusters of methods allowing DBS and CDBS data analysis respectively.
In this paper, however, we focus on the CDBS analysis.

\subsection{Language, Structure and Methods of the Program}
The STACS package was written in python to ease the implementation and its expansion. 
Many of the performance shortcomings of python were mitigated by utilizing efficient external packages such as \verb|NumPy|\footnote{\url{https://numpy.org/}}  and \verb|SciPy|\footnote{\url{https://scipy.org/}}. This also makes STACS compatible with most operating systems.
The program structure is intentionally kept modular to facilitate improvements and replacement of program parts if needed. Currently STACS functions mainly on a command-line interface, with a simple text-prompt-based interface. Alternatively, all components of the program can simply be used in a python environment for more in-depth control. Most python methods have a comprehensive doc-string documentation.

The measurement files are imported into a custom class object. There is a collection of methods to perform miscellaneous calculations on the recorded data, for example for energy calibration, measurement quality assessment, and a comprehensive collection of plotting functions utilizing \verb|matplotlib|\footnote{\url{https://matplotlib.org/}}.

Data recorded from CDBS is imported from a text or xml file into the program. Usually these files contain environmental information about the measurement parameters such as temperature, positron beam position or acceleration voltage, but also information about the detector performance, such as count rate, dead time, etc. For this reason, it was deemed efficient to create a custom class object for Doppler broadening data which contain several attributes such as measurement type and other parameters relevant for the program.
The main objects are two sub-classes, one contains all single detector spectra with additional detector specific parameters. The other contains the CDB spectra as \verb|NumPy| arrays. From this point almost all methods perform tasks on these objects.

\subsection{Evaluation Tools of CDB Spetra}
In the following, we present the evaluation of CDB spectra step by step beginning with the definition and extraction of the ROI, binning of the energy scale, background subtraction and the generation of 1D-projections for the calculation of ratio curves.
Finally, we demonstrate how to determine the effective energy resolution and how to obtain information on the electron binding energy from the 2D-CDB spectra.

\subsubsection{Region of Interest}
\label{sec:roi}
\begin{figure}
    \centering
    \includegraphics[width=.8\linewidth]{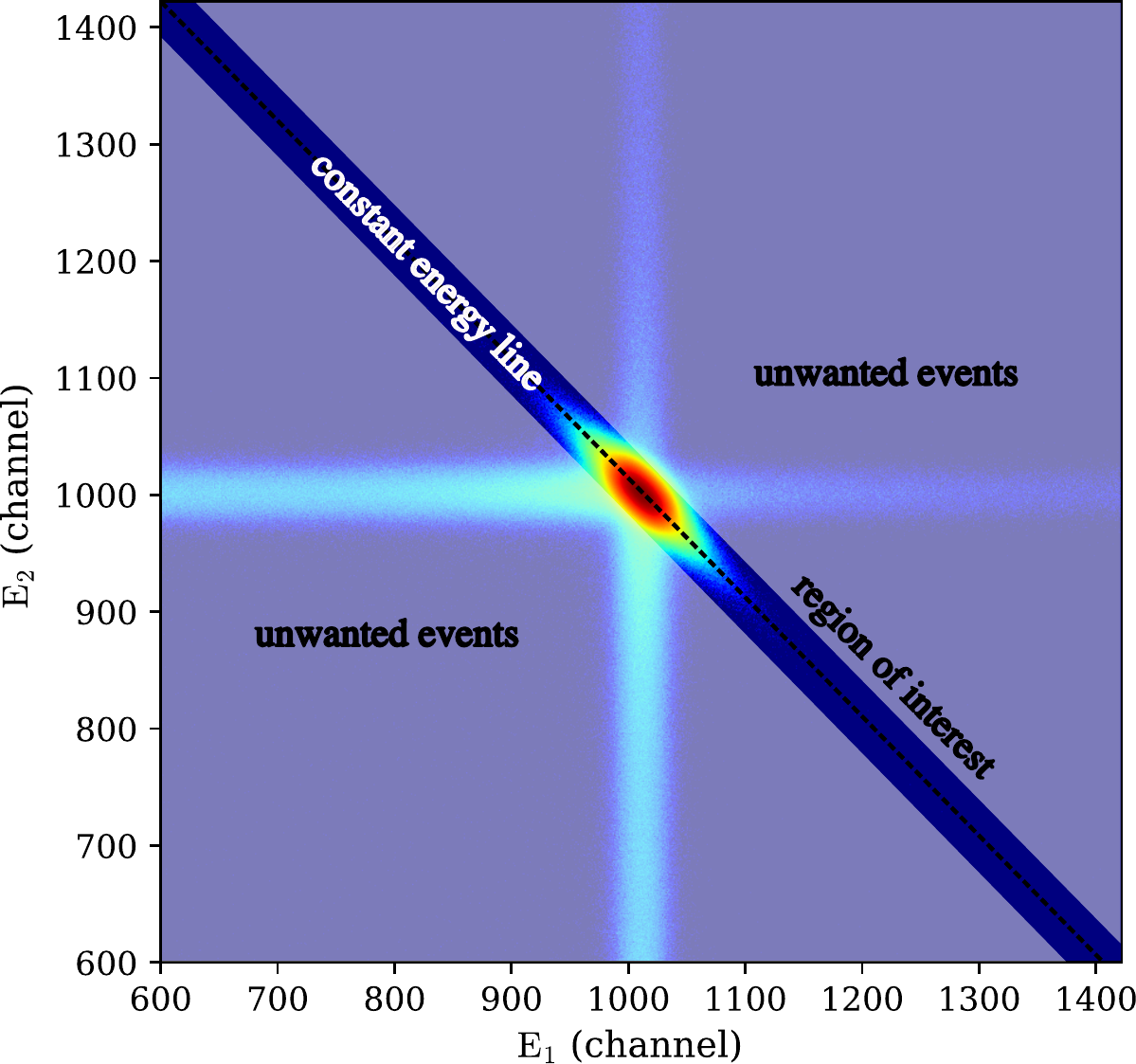}
    \caption{2D histogram of a CDBS measurement showing the location and shape of the region of interest (ROI), parallel to the constant energy line (CEL). Most of the unwanted pileup and Compton scattered events lie outside of the ROI.}
    \label{fig:ROI}
\end{figure}

The  ROI is defined by a simple rectangle centered on the CEL with two relevant parameters, the width orthogonal to the CEL and the tilt with respect to the energy axes of the 2D histogram (see Figure\,\ref{fig:ROI}).
By assuming the 2D histogram to be a coordinate system, the tilt of the CEL can be considered as a slope with $s = -\frac{\Delta e_\mathrm{x}}{\Delta e_\mathrm{y}}$ with $\Delta e_\mathrm{x}$ and $\Delta e_\mathrm{y}$ being the channel width of the detector on the $x$ and $y$ axis respectively.
Hence, if the channel width of both detectors is equal then the slope will be exactly $-1$. 
 
\begin{figure}
    \centering
    \includegraphics[width=0.8\linewidth]{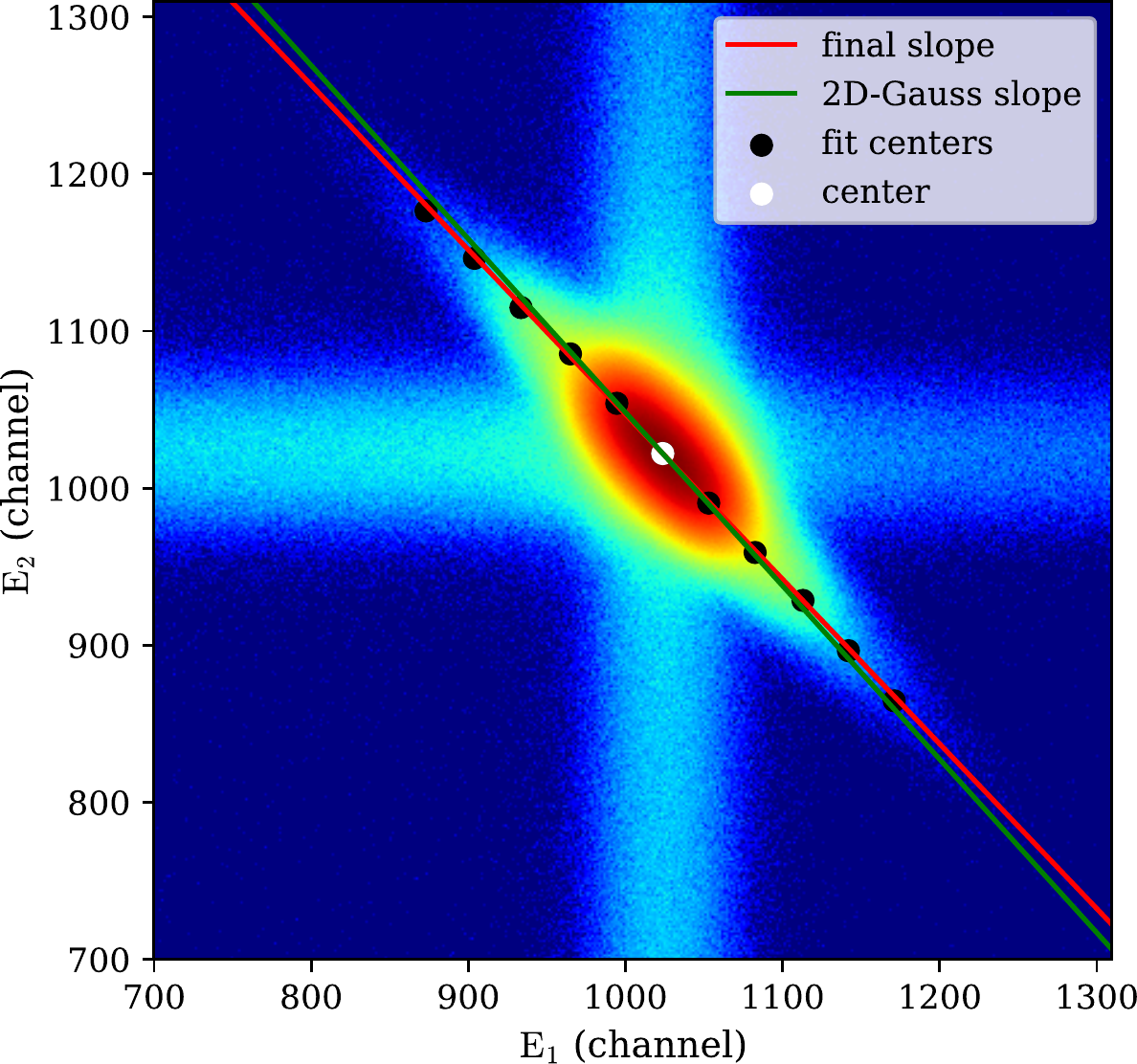}
    \caption{Finding the slope of the CEL: The center (white dot) is determined by fitting a 2D-Gauss into the center region of the spectrum. The according slope (green line) is very sensitive to high count rate artefacts near the center. Afterwards 1D-Gauss fits are done at equidistant points from the center with a slope of $1$. The center of each fit (black dots) is then used for a linear regression fit (red line) defining the slope of the ROI.}
    \label{fig:slope}
\end{figure}

The channel width, $\Delta e$, of the detectors depends on readout parameters such as gain, energy calibration, channel offset, etc.
These readout parameters are hardware specific and will be set beforehand. 
Over longer operation times the energy calibration parameters of the detectors may drift. High count rates can also lead to slightly varying calibration parameters of the detectors. Since it is not practical to do the energy calibration for every measurement anew, the slope is recalculated every time the ROI is determined, as shown in Figure\,\ref{fig:slope}. From a computational point of view this is not an issue, since the entire ROI positioning process is very fast (less than $ 1 \  \mathrm{s}$).

For the ROI location the center of the spectrum has to be determined first. For this purpose a 2D-Gaussian fit is performed in the center of the 2D histogram. In principle, the tilt and center of the 2D-Gauss should already determine all the parameters for the location of the ROI.
The tilt obtained by the 2D-Gauss fit, however, is usually inaccurate due to minor differences in energy resolution and response behavior of the two detectors in the high intensity center region of the spectrum.
As shown in Figure\,\ref{fig:slope}, the slope of the 2D-Gauss fit naturally is more sensitive to the center of the CEL where the count rate is the highest. 
As such, the slope given by the 2D-Gauss is sensitive to disturbances of the spectrum in the center region where high count rates can cause the CEL to "smear" out along one of the detector axes creating a small deviation of the 2D-Gauss slope. Since this effect does not necessarily harm the quality of the measurement, it is to be ignored for the determination of the CEL slope. 
However, in order to reliably determine the actual slope of the ROI several Gauss fits are preformed perpendicular to the CEL. The locations for these fits are distributed evenly along the CEL and  chosen such that they cover most of the CEL. This method gives a very accurate estimation of the slope of the CEL because large Doppler shift regions are sampled as well. 
A large deviation between the 2D-Gauss and the actual calculated slope, however, can be an indication of detector faults  or very high or very low count rates and should be taken into account when proceeding with the data evaluation.

Once the slope of the ROI has been determined, its width perpendicular to the CEL has to be calculated. Generally the ROI should be as wide as possible in order to maximize the amount of data included for evaluation. At a certain width though the Compton and pileup background will start to influence the projection significantly. Therefore, the ROI width is set to one FWHM of the coincidence energy resolution, (a detailed analysis of the ROI width is given in the appendix, Section\,\ref{sec:app}). The ROI length along the CEL is not relevant for now, since it is essentially determined by the number of energy bins, which are considered in the next step.

When position and slope have been determined the ROI will be cut out from the rest of the spectrum that separates it from the background and greatly increases the speed of the binning process. 
In order to accurately cut out the ROI a custom algorithm is used. 
This process is essentially an image processing exercise where one point in the 2D histogram corresponds to a pixel in an image. 
Subsequently, an aliasing problem has to be considered since the section to be cut does not line up with pixel boundaries, i.e., the cut line will pass through some of the  pixels at the boundaries of the ROI. 
The simplest solution of this problem is a binary algorithm: If the center of the pixel is inside the boundary region the entire pixel is counted, otherwise it is ignored. However, as this same procedure will be applied to perform the binning of the ROI afterwards, a more accurate method is required. Instead, the area fraction of the pixel inside of the ROI will be calculated, and the value of the pixel will be included proportionally to that area fraction. Since the slope of both the ROI and the bin boundaries is always close to $\pm 1$ only the distance of the pixel center from the boundary line is required in order to determine the proportion of the pixel that lies inside the boundary. If the width of one pixel is set to 1, the area fraction of the pixel inside the boundary region $\sigma$ can be calculated:
\begin{equation}
    \label{eqn:px_d}
    \sigma = \left.
    \begin{cases}
        1, & \text{for} \ d \geq \frac{\sqrt{2}}{2} \\
        0.5 + d\left( \sqrt{2} - |d|\right), & \text{for} \ \left| d \right| \leq \frac{\sqrt{2}}{2} \\
        0, & \text{for} \ d \leq  -\frac{\sqrt{2}}{2}
    \end{cases}
    \right.
\end{equation}
where $d$ is the distance of the pixel center to the boundary. If the center is outside the boundary region $d$ becomes negative. 
Since this approach is based on partially pixel-wise calculations it is time consuming. The algorithm has been optimized by the utilization of \verb|NumPy| arrays.
In this context, \textit{partially pixel-wise}  means that not all pixels of the 2D histogram have to be considered. Most pixels are either entirely inside or outside of the ROI and only those being intersected with the boundaries will be treated by the algorithm.

\subsubsection{Binning}
\label{sec:bin}
Once the ROI has been separated from the histogram most of the unwanted detected events have been eliminated. 
The data, however, is still contained in a 2D array. For projection, one could simply rotate the array and project it onto one axis. This approach, however, yields two problems: a rotation always induces interpolation errors as the data is discrete in two dimensions. Additionally, at high Doppler shift energies the data becomes increasingly noisy due to the low number of counts. As a solution, the data is split into energy bins. These bins can be altered depending on the statistical quality of the measurement and are generally just given as a list of energies starting with $511 \ \mathrm{keV}$.

\begin{figure}[h]
    \centering
    \includegraphics[width=.88\linewidth]{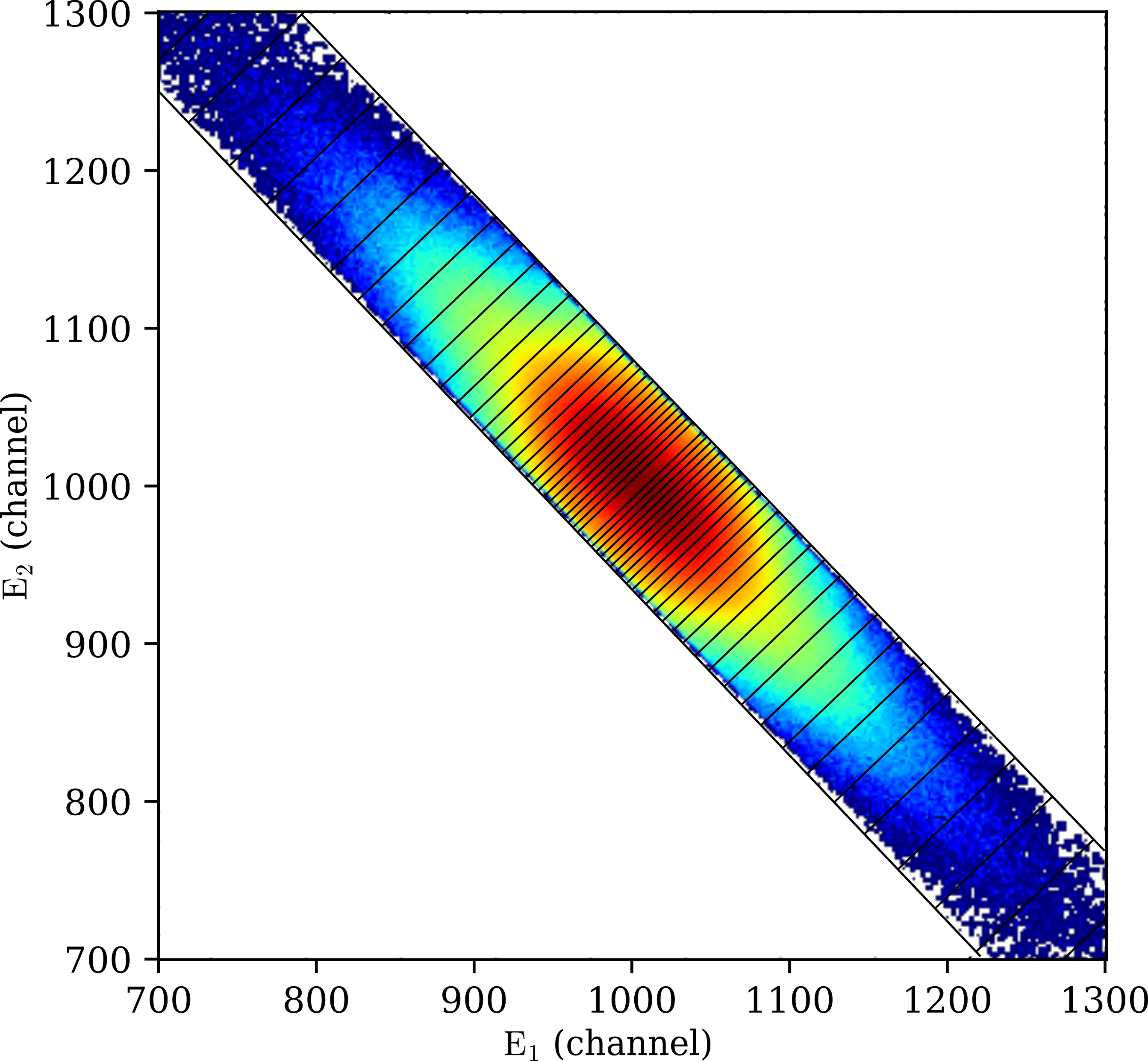}
    \caption{The 2D histogram with the ROI separated from the background. The bins (black lines) linearly increase in size to compensate for less count rate in the outer regions of the ROI.}
    \label{fig:bin}
\end{figure}

By using the same algorithm as for cutting out the ROI (Equation \ref{eqn:px_d}), the binning algorithm now cuts the ROI into the corresponding energy bins as illustrated in Figure\,\ref{fig:bin}.
At this point the energy calibration of the detectors is required in order to correctly determine the bin positions in accordance with the actual energies of the annihilation $\gamma$-quanta. Note that only the parameters of one of the detectors are needed as the tilt of the ROI correlates the relevant parameters of the detectors with each other. By default STACS uses the energy calibration of detector 2 from Figure\,\ref{fig:bin}. 
The lines that separate the bins can again be considered as linear equations with a y-intercept at a corresponding detector 2 channel. Hence, the desired binning is converted into y-intercept values of the detector 2 axis with a slope perpendicular to the ROI slope calculated in Section\,\ref{sec:roi}. All pixels in each bin will now simply be summed up. Over all, the binning process is by far the computationally most intensive, as a large portion of the pixels lying at  the bin boundaries are evaluated in the cutting process, requiring usually a few seconds per spectrum. This, however, obviously also depends on the number of energy bins and the single core performance of the processor.

\subsubsection{Background Subtraction}
\begin{figure}
    \centering
    \includegraphics[width=.85\linewidth]{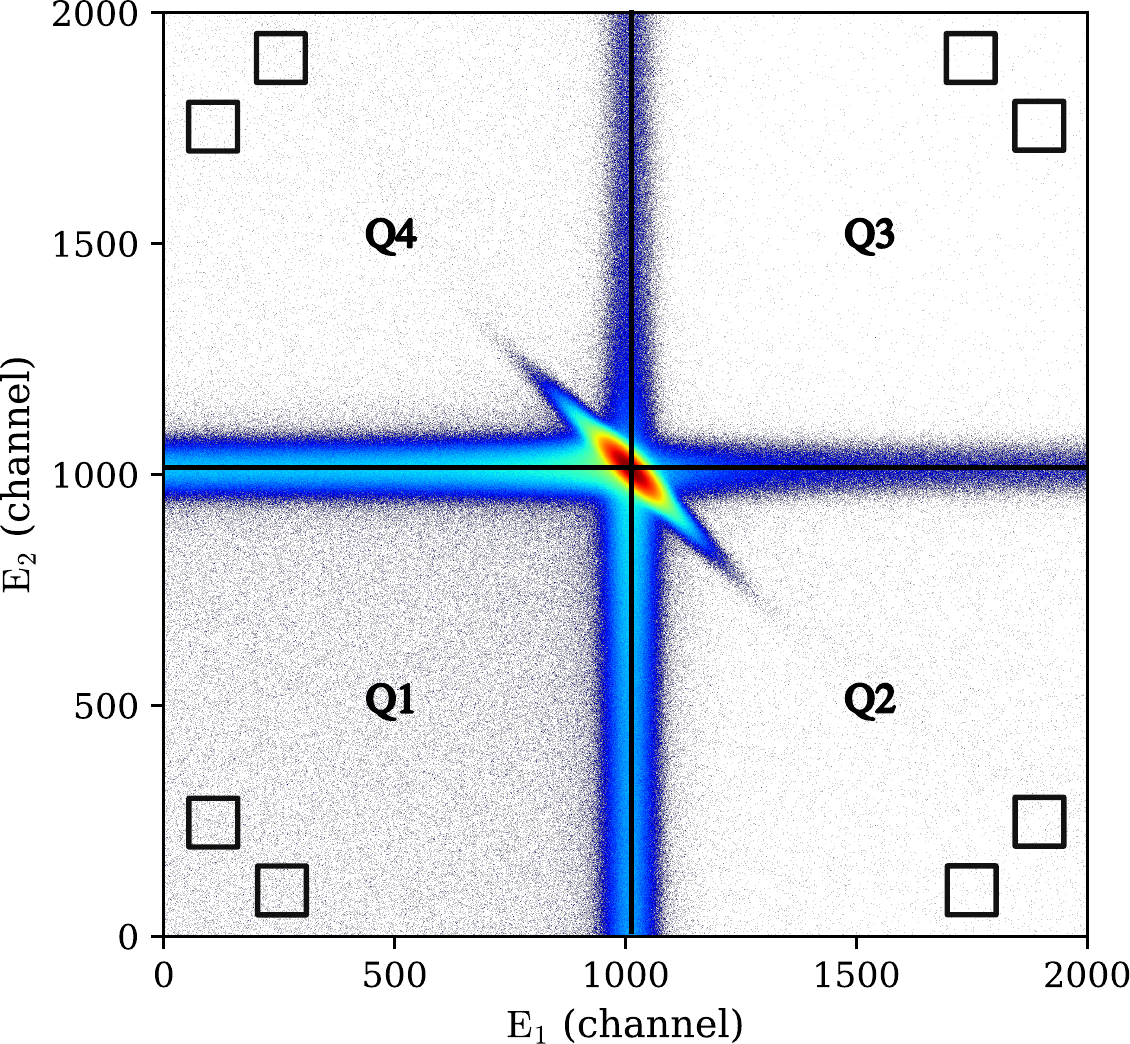}
    \caption{2D histogram with the ROIs for background subtraction. The average background is calculated for each quadrant of the histogram. The regions are chosen in a way where they are not influenced by the annihilation line. In order to subtract the background from the annihilation line only quadrants Q2 and Q4 are used, since the annihilation line only extends into those regions.}
    \label{fig:bg_roi}
\end{figure}
A large amount of the unwanted background and erroneous counts have already been eliminated with the regular process described so far. 
However, the entire coincidence spectrum also has a random constant background which is difficult to distinguish from actual counts at very high Doppler shifts. 
The magnitude of this remaining background is determined by defining four ROIs, two in the top left  and two in the bottom right of the 2D histogram (see Figure\,\ref{fig:bg_roi}).
\begin{figure}[h]
    \centering
    \includegraphics[width=\linewidth]{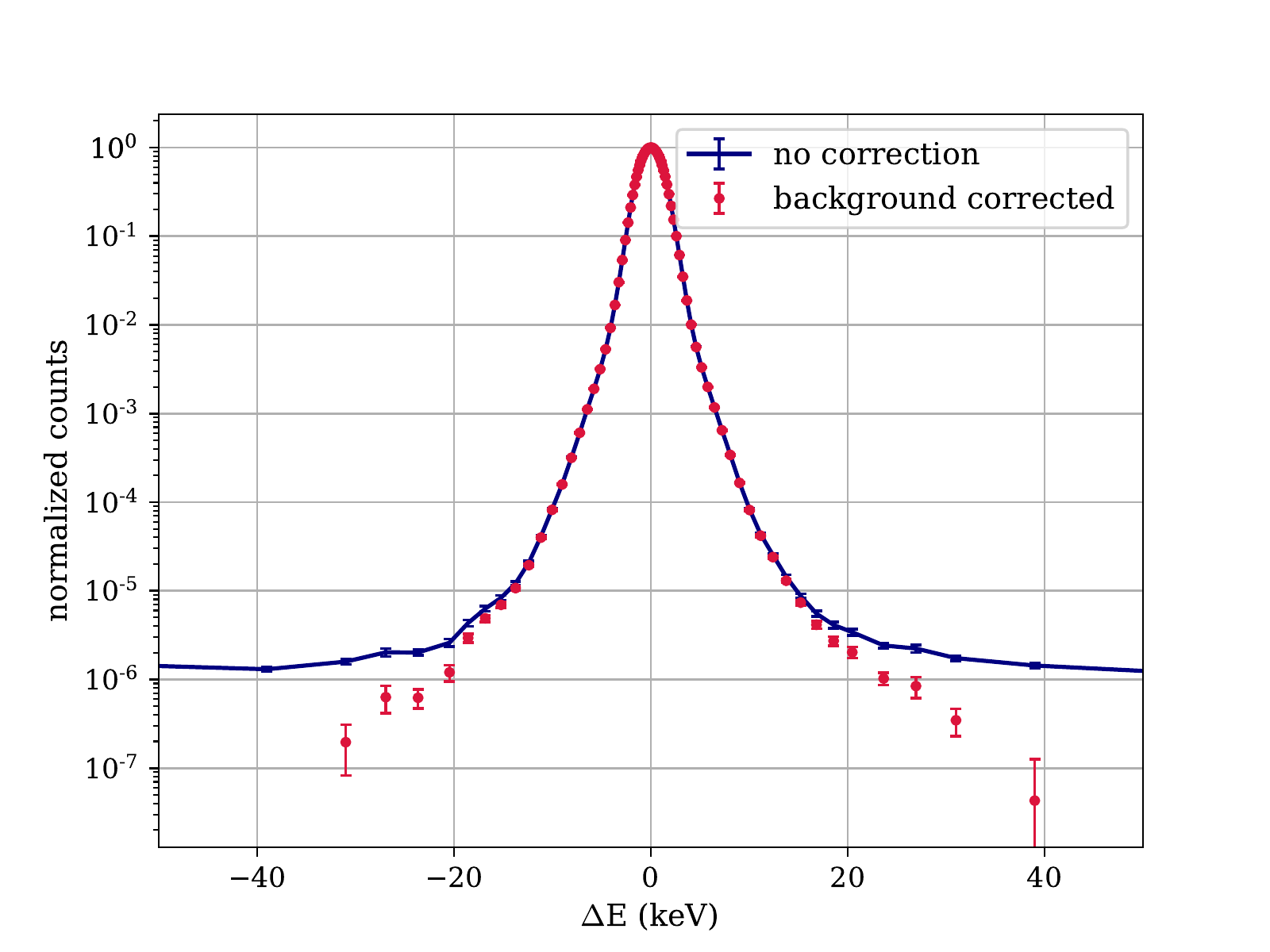}
    \caption{Projection of the 511\,keV photo peak without (blue) and with (red) background subtraction. At Doppler shifts greater than $40 \ \mathrm{keV}$ in the uncorrected projection a constant background caused by random coincidences remains which vanishes by the background correction.
        }
    \label{fig:bg_sub_proj}
\end{figure}
Within these ROIs the average counts per pixel are determined to create a separate array, where every pixel just contains the average background value. This background array is then treated analogously to the coincidence data including the binning. The resulting average background per energy bin of the projection is now subtracted from the projection data itself. An example of the efficacy of this method for background correction is shown in Figure\,\ref{fig:bg_sub_proj}; note the large region of the energy scale reaching up to very large Doppler shift where the background is further reduced by about one order of magnitude. In the appendix (Section\,\ref{sec:app}) we additionally verify the effectiveness of this background subtraction method.

\subsubsection{Energy Resolution at CDBS}
\label{sec:eres}
\begin{figure}[h]
    \centering
    \includegraphics[width=\linewidth]{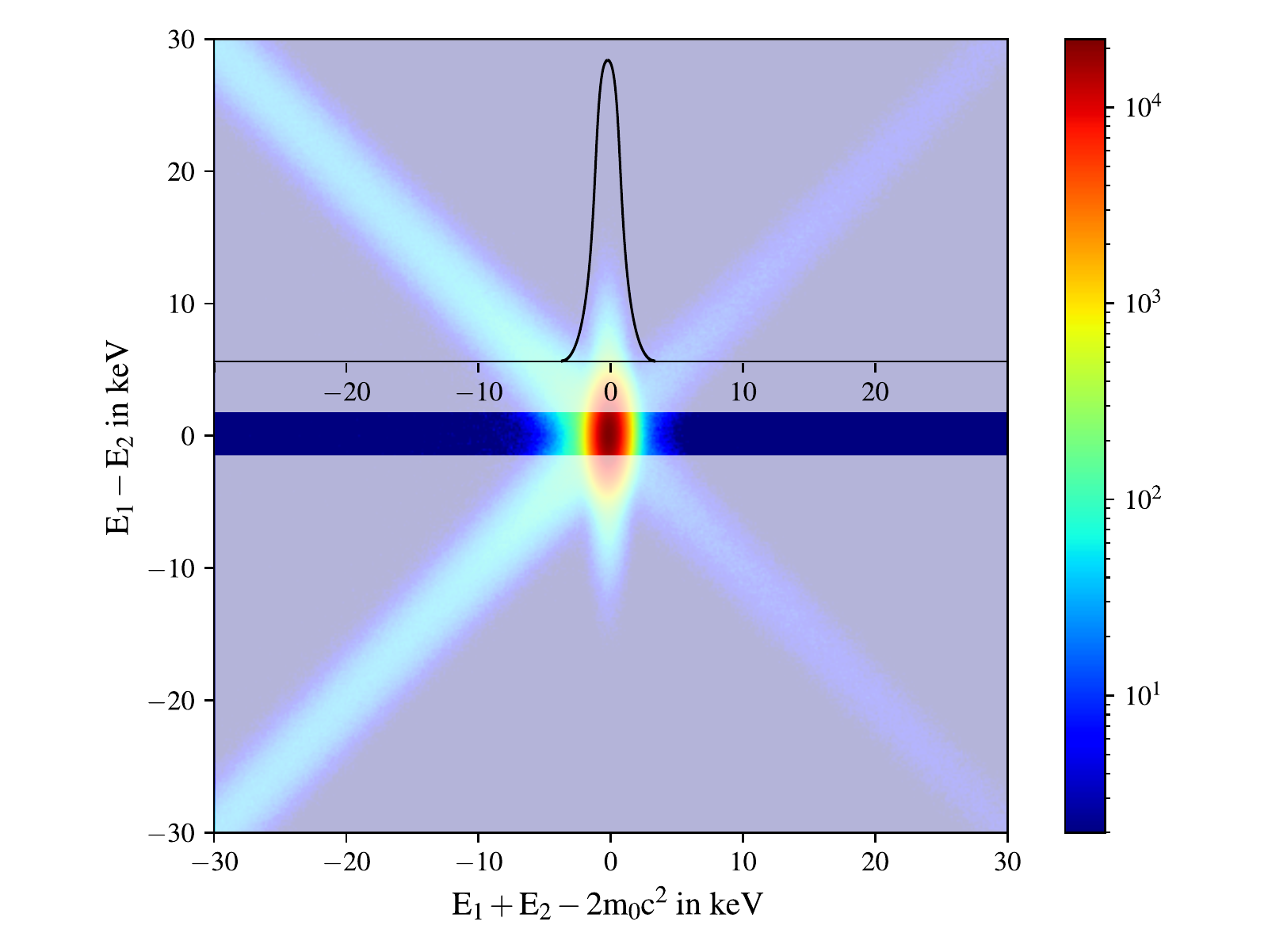}
    \caption{Determination of  the energy resolution of a CDBS measurement at $511 \ \mathrm{keV}$. 
    The 2D histogram is rotated such that the $1022 \ \mathrm{keV}$ CEL is vertical. A Gauss fit on a projected slice around the center of the CEL yields the effective energy resolution.
    }
    \label{fig:eres_coinc}
\end{figure}

In theory, according to Poisson statistics the effective energy resolution of CDBS is expected to be a factor of $\sqrt 2$ higher than in conventional DBS since both detectors contribute to the measured 511\,keV annihilation line equally.
In the 2D histogram, the energy resolution of the respective single detector can be obtained by the spectrum projection along the x and y axis.
For the determination of the ROI width, however, it is imperative to know the effective energy resolution of the CDBS measurement.
The energy resolution at the photo peak is determined by projecting along the CEL and performing a Gauss fit, as shown in Figure\,\ref{fig:eres_coinc}.
First, the 2D histogram is rotated such that the $1022 \ \mathrm{keV}$ CEL is vertical. 
Then the effecitve energy resolution at 511\,keV is determined by performing a Gauss fit only at a projected slice around the center of the CEL in order to suppress effects from the Compton and pileup background.

\subsubsection{Binding Energy}
So far it was assumed that all of the annihilation events lie on the $1022 \ \mathrm{keV}$ CEL. 
This is, however, not exactly the case as the total released annihilation energy $E_\mathrm{an}$ is reduced by the binding energy of the annihilating electron $E_\mathrm{bind}$:
\begin{equation}
    E_\mathrm{an} = \frac{1}{2} (m_\mathrm{e}c^2 - \Delta E + m_\mathrm{e}c^2 + \Delta E) - E_\mathrm{bind}
\end{equation}
Note that the binding energy of the thermalized positron can still be assumed to be negligible. 
This means that the measured energy of the annihilation events from each electron orbital are situated on a different CEL. Usually, for defect spectroscopy and material studies using CDBS the observation of this effect is not relevant, as the electron binding energy is very small compared to the electron mass. 
However, it is possible to detect the influence of tightly bound electrons in a CDB spectrum in this way, although even the improved energy resolution at CDBS is limiting this effect. 
In conjunction with a ratio curve, extracting this information might be useful though, as it can increase specificity \cite{Pikart2013, Cizek2010}.
In order to extract the binding energy shift the CDB spectrum is evaluated using multiple very thin ROIs effectively binning the spectrum in small slices parallel to the CELs. 

\subsubsection{Projections and Ratio Curves}
\label{sec:ratio}
The spectrum obtained by the projection of the annihilation line contains information about the electron momentum distribution of the sample. 
For a pure element the annihilation line at Doppler shifts greater than the Fermi energy can be described by a superposition of all electron states weighted by their core annihilation probabilities. 
For more complex materials, such as alloys, the contribution of different elements to the CDB spectra is often visualized by so-called ratio curves.
For this purpose, STACS includes functionality to easily produce such ratio curves from CDB spectra. We offer a lot of flexibility here; since the CDB spectrometer at NEPOMUC, which is the facility that STACS has been primarily developed for, has a total of ten detectors, STACS is able to produce ratio curves using different detector pairs. Additionally, STACS can average detector pairs to produce statistics improved data. Note that this requires careful calibration of the detectors since the energy resolution can have a large influence on the shape of ratio curves.

\begin{figure}[h]
    \centering
    \includegraphics[width=1\linewidth]{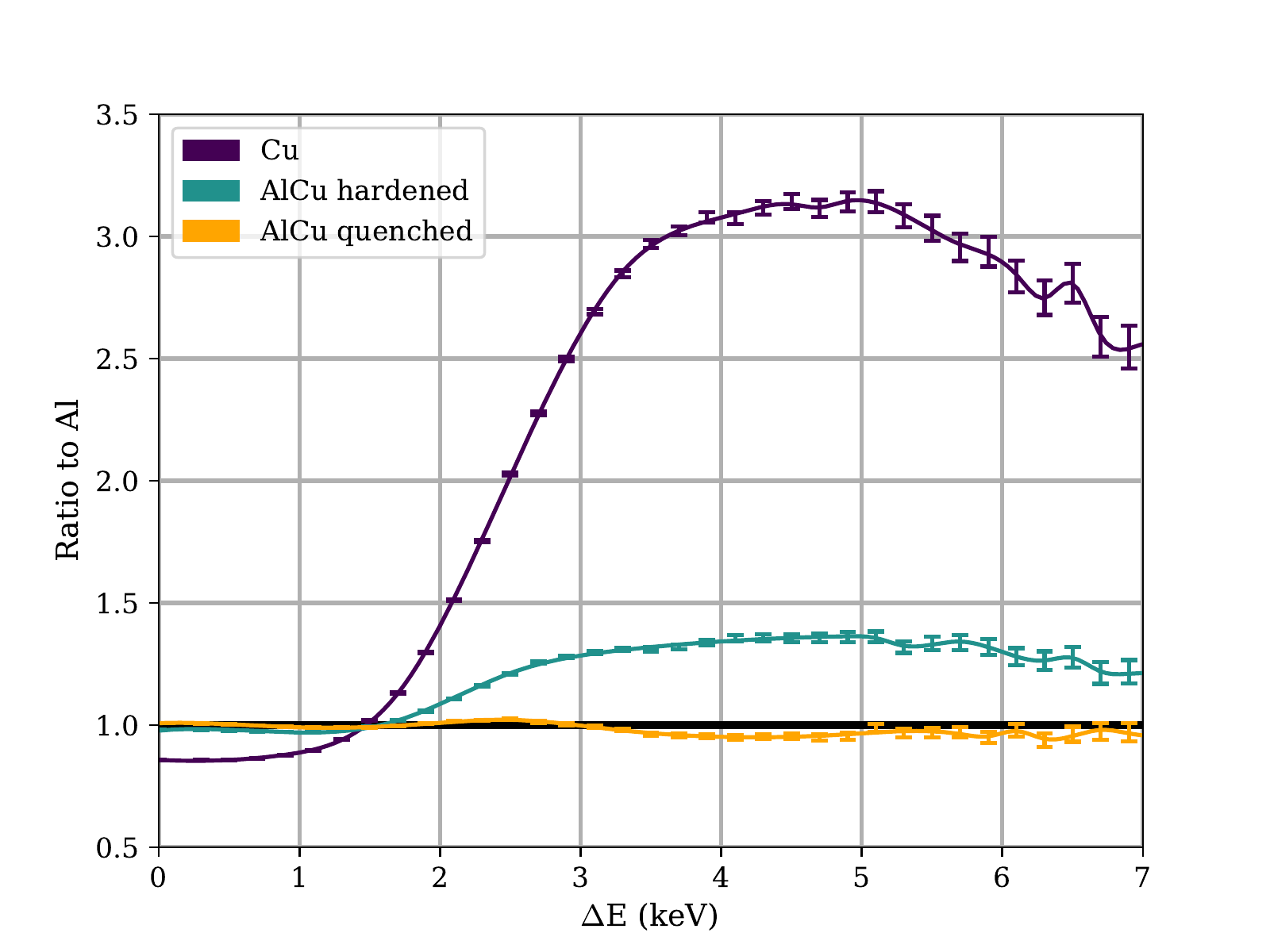}
    \caption{Ratio curves of a precipitation hardened AlCu alloy and a Cu reference with respect to annealed Al. Since Cu precipitates in the age-hardened alloy trap positrons the ratio curve differs from pure Al and shows a distinct Cu signature. After solution annealing and subsequent quenching only the confinement peak at around $2.3 \ \mathrm{keV}$ is slightly visible resulting from positrons trapped in vacancies.}
    \label{fig:ratio}
\end{figure}

In order to demonstrate the functionality of STACS we analyse different CDB spectra recorded for pure Al and Cu as well as for samples of an age-hardened AlCu alloy.
For the comparison of the spectra of age-hardened and quenched AlCu samples, the ratio curves are shown with respect to annealed Al in Figure\,\ref{fig:ratio}; the ratio curve of Cu serves as reference. 
As expected the hardened AlCu alloy shows a distinct Cu signature
since the presence of Cu precipitates, exhibiting higher positron affinity than the Al host lattice, leads to positron trapping. After solution annealing and quenching the precipitates are dissolved. Since single atoms do not form attractive potentials strong enough for trapping positrons the Cu signature in the CDB spectra vanishes. 
The formation of vacancies, however, lead to positron trapping that results in the so-called confinement peak visible at around $2.3 \ \mathrm{keV}$.

\section{Summary}
The goal of STACS was to create a software solution that is able to take in CDBS data in the form of 2D histograms and analyze the spectra all the way, including the creation of ratio curves. Using python provides several advantages; this allows the software to be computationally efficient by using external open source packages, but it also makes the software easy to use and even expand upon. STACS can be easily imported into existing data analysis scripts, which are commonly already written using python. Most importantly, STACS is able to efficiently extract the projection of the annihilation line from CDBS histograms without introducing data processing artefacts. The accurate subtraction of a constant background provides a very effective and simple way of significantly improving the signal-to-noise ratio, even compared to more complex background subtraction schemes.
For the first time it is possible to combine data from multiple detector pairs and measurement runs at differing points during data processing, reducing the measurement time required.  
In addition to the pure data extraction STACS includes several further functions that make trouble shooting and in depth data analysis simple. 

For the scientific community it is most important that we offer the software as open source package which can be downloaded for free. STACS is licensed using the GNU GPLv3 license and can be downloaded \href{https://gitlab.lrz.de/tum-frm2-positrons/stacs}{here}\footnote{\url{https://gitlab.lrz.de/tum-frm2-positrons/stacs}}. Users are invited to apply, modify and adapt STACS by referring to this publication. We also encourage users to raise potential issues and to submit corrections or additions as merge requests.

\section{Acknowledgement}
Financial support by the German federal ministry of education and research (BMBF) within the Project No. 05K22WO7 is gratefully acknowledged.

\bibliography{main.bib}
\section{Appendix}
\label{sec:app}
\subsection{Determination of the Optimal ROI Width}
\begin{figure}[h]
	\centering
	\includegraphics[width=0.93\linewidth]{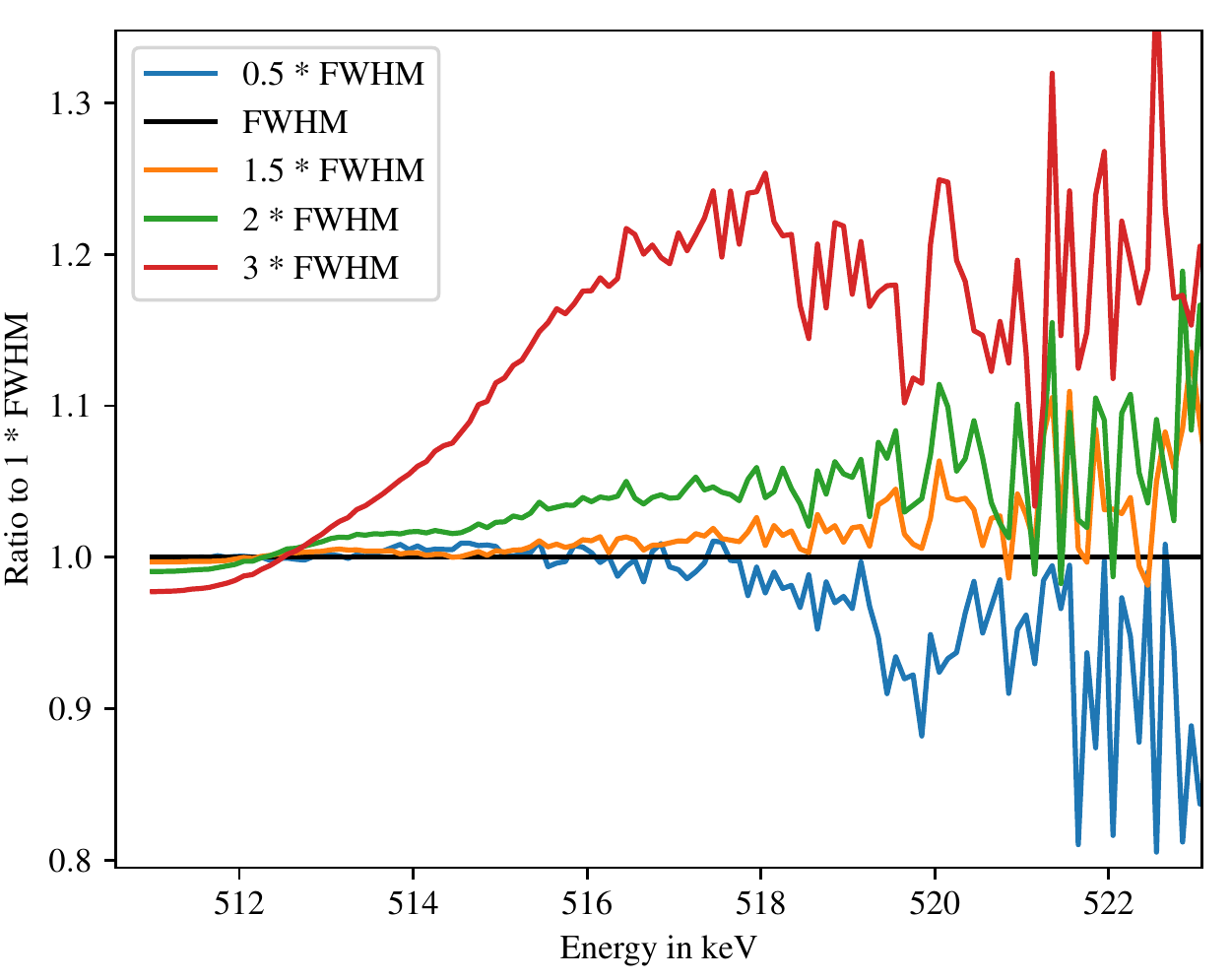}
	\caption{Projections at different ROI widths and their respective ratio to a width of $1 \cdot \mathrm{FWHM}$: At larger ROI widths the broadening of the peak and consequently the ratio compared to $1 \cdot \mathrm{FWHM}$ increases because more counts from the Compton and pileup background lie inside the ROI.}
	\label{fig:roi_w}
\end{figure}

To determine the optimal ROI width the FWHM of the coincidence energy resolution was chosen as a reference. Ratio curves were produced by increasing and decreasing the ROI width compared to 1 FWHM, the results can be seen in Figure\,\ref{fig:roi_w}. It can be seen, that for increasing ROI width settings a significant amount of counts originating from the pile up and Compton regions in the 2D histograms contributes to a higher ratio at increasing Doppler shifts. Reducing the width of the ROI below 1 FWHM does not significantly influence the ratio. Since a smaller region of interest always represents counting less events and hence reducing the statistics of the measurements, the ROI is by default set to 1 FWHM in STACS as this represents a good compromise.

\subsection{Analysis of the Background Subtraction}

\begin{figure}[h]
	\centering
	\includegraphics[width=0.95\linewidth]{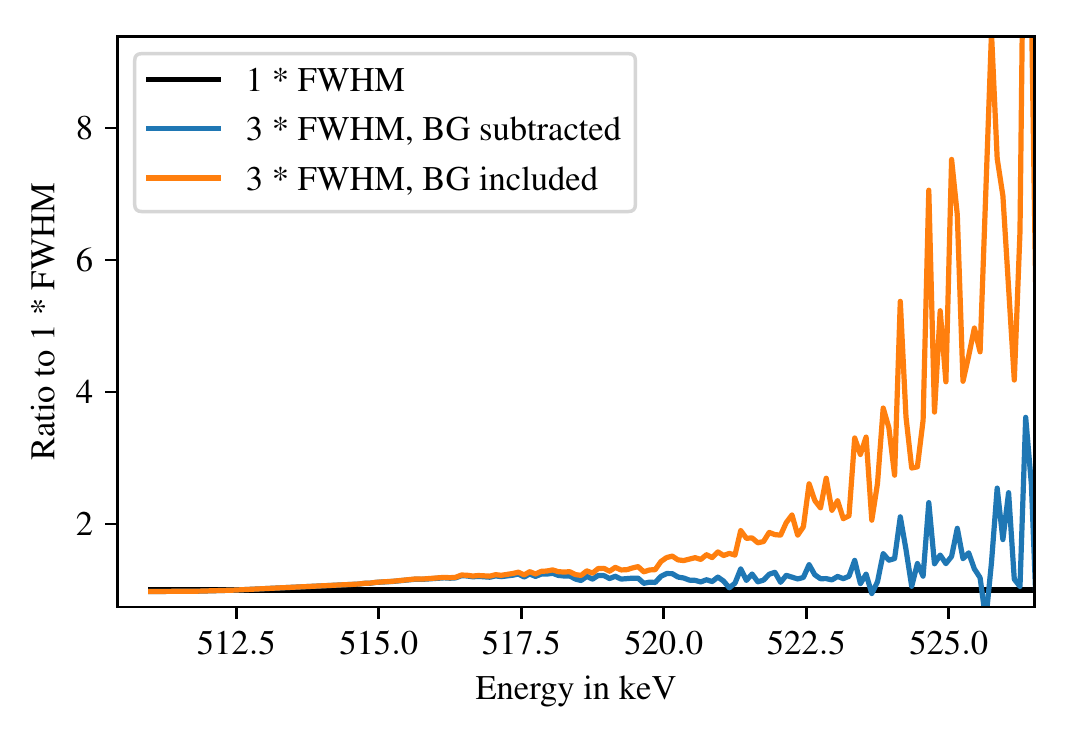}
	\caption{In this example the comparison of ROI width from Figure\,\ref{fig:roi_w} is done once with and without background subtraction. This example is used, since a wider ROI inevitably leads to more background events being included. This can be seen clearly starting at approximately $517 \mathrm{keV}$. As the count rate at high Doppler shifts gets low this effect starts to dominate. When the background is subtracted this effect is successfully negated and the actual influence of the ROI width on the projection can be analyzed and one can see that at high Doppler shifts the effect of the ROI width decreases.}
	\label{fig:bg_sub}
\end{figure}

One way the viability of our background subtraction method can be analyzed is by comparing the effect of different ROI widths. As mentioned before, when the region of interest is widened a major effect on the projection comes from the pile up and Compton events. However, at high Doppler shifts the background will also introduce increased counts when the ROI width is increased, since more background events fall within the ROI. In Figure\,\ref{fig:bg_sub} we demonstrate that the effect of the background at high Doppler shifts is effectively mitigated by the background subtraction.

\end{document}